\documentclass[preprint,12pt,authoryear]{elsarticle}

\usepackage{amssymb}

\usepackage{paralist}
\usepackage{soul}
\usepackage{booktabs}
\usepackage{tikz}
\usetikzlibrary{shapes,arrows,calc,backgrounds}
\usepackage{caption}
\usepackage{subcaption}
\usepackage{enumitem}
\usepackage{appendix}
\usepackage{multirow}
\usepackage{lscape}
\usepackage{tabularx} 
\usepackage{longtable}

\DeclareCaptionLabelFormat{AppendixATables}{Table A#2}

\journal{IJHCS}

\begin{document}

\begin{frontmatter}

\title{A Critical Reflection on the Use of Toxicity Detection Algorithms in Proactive Content Moderation Systems}

\author[add1]{Mark Warner\corref{cor1}}
\ead{mark.warner@ucl.ac.uk}

\author[add2]{Angelika Strohmayer}
\ead{angelika.strohmayer@northumbria.ac.uk}

\author[add3]{Matthew Higgs}
\ead{m@higgs.ac}

\author[add4]{Lynne Coventry}
\ead{l.coventry@abertay.ac.uk}

\cortext[cor1]{Corresponding author}
\address[add1]{University College London, London, United Kingdom}
\address[add2]{Northumbria University, Newcastle-upon-tyne, United Kingdom}
\address[add3]{Independent Researcher,  United Kingdom}
\address[add4]{Abertay University, Dundee, United Kingdom}



\begin{abstract}

Toxicity detection algorithms, originally designed for reactive content moderation systems, are being deployed into proactive end-user interventions to moderate content. Yet, there has been little critique on the use of these algorithms within this moderation paradigm. We conducted design workshops with four stakeholder groups, asking participants to embed a toxicity detection algorithm into an imagined mobile phone keyboard. This allowed us to critically explore how such algorithms could be used to proactively reduce the sending of toxic content. We found contextual factors such as platform culture and affordances, and scales of abuse, impacting on perceptions of toxicity and effectiveness of the system. We identify different types of end-users across a continuum of intention to send toxic messages, from unaware users, to those that are determined and organised. Finally, we highlight the potential for certain end-user groups to misuse these systems to validate their attacks, to gamify hate, and to manipulate algorithmic models to exacerbate harm.

\end{abstract}


\begin{highlights}
\item We describe contextual factors impacting on perceptions of toxicity and system efficacy
\item We describe system efficacy across different end-user groups
\item We identify and describe ways in which these systems are open to abuse
\item We discuss how these systems could have unintended consequence
\item We present considerations for designing proactive moderation systems
\end{highlights}


\end{frontmatter}



\section{Introduction}

As people's communications and interactions move increasingly online, societies have experienced increases in different forms of online harm. In June 2024, 68\% of UK internet users (+13) had been exposed to at least one potential online harm within the preceding 4 weeks of being surveyed. Of those, the most common type of content encountered was hateful, offensive or discriminatory in nature~\citep{Ofcom_report}. Exposure to this type of toxic content can have a significant impact on mental health and wellbeing of victims~\citep{duggan2017online, davidson2019adult,hinduja2010bullying}, including depression and anxiety, as well as suicidal ideation~\citep{stevens2021}. The impact may also be felt by producers of online abuse, such as damage to identity and reputation~\citep{madden2010}. Whilst some online abuse is created by organised groups in coordinated attacks on platforms like Reddit~\citep{mittos2020a,mittos2020b}, 4Chan~\citep{mittos2020a, mittos2020b,mariconti2019}, Gab~\citep{zannettou2018gab}, and Twitter (now `X', though we continue to use `Twitter' throughout this paper since the fieldwork was carried out prior to the name change)~\citep{chatzakou2017hate, chatzakou2017}, other abuse is less organised. For example, prior work finds everyday people posting messages containing profanities, obscenities, and offensive statements when in ``hot states''~\citep{wang2011regretted,sleeper2013}. Similarly,~\cite{cheng2017anyone} found everyday users can engaging in online toxic (or `trolling') behaviours when in certain emotional states and when prompted within certain discussion context.

Several algorithmic approaches for detecting toxic speech have been developed and evaluated (e.g.,~\cite{ketsbaia2020, agrawal2018deep, georgakopoulos2018}) with these and similar algorithms used to support online content moderation on social media platforms. This approach to moderation has the benefit of limiting human exposure to toxic content (i.e., to human moderators~\citep{steiger2021}), yet is fraught with ethical concerns related to fairness, bias, and freedom of speech~\citep{vaccaro2020end}. An alternative approach to using these same or similar algorithms involves proactive moderation at the point of message creation to nudge senders into making better choices~\citep{kiritchenko2021confronting}. Similar nudging approaches have been explored to help protect users against inappropriate disclosures, through privacy nudges~\citep{wang2014field,wang2013privacy}, and to help prevent misinformation~\citep{jahanbakhsh2021exploring}. These proactive approaches help to address concerns related to freedom of speech. They also push the responsibility of dealing with toxic content away from receivers and human moderators, and onto senders. Moreover, they could allow for previously unmoderated spaces (e.g., WhatsApp) to incorporate moderation despite the closed and secure nature of these platforms. Online social networks have started embedding these systems into their platforms~\citep{katsaros2021reconsidering, techradar2021} and research has started to explore the role of interaction design for delivering reflective prompts~\citep{royen2017, jones2012reflective}. While we have seen some deployments of algorithmic-driven proactive content moderation systems within platforms, there has been little in the way of critique of these systems. Critique is important for us to garner a more holistic understanding of not only the opportunities but also limitations of these types of systems. In turn, this is important because these proactive rather than reactive systems are integrated into the front-end of communication platforms and act to mediate communications with the use of AI~\citep{hancock2020ai}. To start developing a more critical understanding of these systems, we pose the following research questions: 

\begin{itemize}
    \item[RQ1] What are the broader challenges of embedding toxicity-detection algorithms into socio-technical systems such as proactive moderation interventions?
    \item[RQ2] How might different users respond to proactive forms of moderation?
    \item[RQ3] What are the potential risks of algorithmic-driven proactive content moderation systems?
\end{itemize}




To address these questions, we ran a series of design workshops in which participants were asked to design an embedded proactive content moderation intervention into an imagined mobile phone keyboard. The intervention would respond to toxic messages at the point of writing with the aim of reducing levels of toxicity within messages. These workshops were conducted with distinct stakeholder groups ($N=18$): (1) five early career HCI researchers due to their expertise in interaction design and knowledge of HCI (2) four academic and non-academic experts in cyberbullying and online hate (including an expert from a UK-based hate charity)  to highlight knowledge of professionals working in the sector, (3) five online content moderators to ensure we embed knowledge from those currently working to reduce toxicity within our research, and (4) five end-users (i.e., people who have previously sent toxic content which they later regretted sending) to include those who's behaviour may be changed with these intervention designs. 

Our findings highlight the complexity of integrating toxicity algorithms into proactive moderation systems. Prior work by~\cite{caplan2018content} and~\cite{gillespie2018custodians} highlights the difficulty of incorporating contextual factors into content moderation. Previous research has also identified specific contextual factors that may impact on moderation decisions~\citep{Jhaver2018,schoenebeck2021drawing} and how identified factors can impact on severity of harm~\citep{scheuerman2021framework}. We draw on this prior work to help us understand the role of context within a proactive moderation paradigm, and build on this work by contributing a holistic and user-centred understanding of the different  contextual factors around the proactive evaluation of toxicity within messages, which range from understanding the relationship between conversation partners, to social histories of oppression and power structures. Expanding on prior work by~\cite{cheng2017anyone} we explore end-users for which such an intervention may be effective and why, allowing us to identify who we are designing for. We also build on previous work conducted within the reactive moderation paradigm by providing support for these proactive moderation systems to be circumvented by users to avoid detection~\citep{chancellor2016thyghgapp,feuston2020conformity, gerrard2018beyond,jhaver2019human}. Our work expands on prior findings related to system misuse by contributing new understanding into how proactive moderation systems could be used to validate and gamify hate and be used to manipulate the algorithmic models to exacerbate harm. Drawing on our findings and those from prior research, we discuss considerations for designing and developing these moderation systems, suggesting the need for improved interdisciplinary work to address the contextual complexities that could result in inequalities, as well as harms through misuse of these systems.

\section{Related work}
Proactive moderation systems are user-facing interventions that rely on users responding to prompts within the system. As such, we first explore prior research looking at why people post toxic content online. As our research explores the design of systems that embed toxicity detection algorithms, we draw on both Machine Learning (ML) and Natural Language Processing (NLP) literature, focusing in particular on the role of context, and the embedding of inequality within toxicity detection models. We then explore HCI and design work related to interfaces designed to promote reflection around the creation of toxic online content. Finally, as our work takes a critical perspective to highlight the potential unintended consequences of these systems, we explore literature on the abuse of algorithmic systems, and how algorithms can be used for multiple purposes (both good, and bad). 

\subsection{Why do people post toxic content online?}
Online toxicity\footnote{Harmful speech can include hateful, offensive, and abusive language, among others. To our knowledge, there is no standard definition that encapsulates these differences~\citep{waseem2017, grondahl2018, davidson2017}. As such, we use the term ``toxic"  to refer collectively to these different forms. Where work has explored a specific type of speech, we will use that specific term. } can be organised and coordinated (see:~\cite{mittos2020a,mittos2020b,mariconti2019, zannettou2018gab, chatzakou2017hate, chatzakou2017}), with people acting alone or in groups~\citep{erjavec2012you}. Those with a desire for social recognition or with heightened moral arousal towards the target and their behaviour, may be more likely to participate in online firestorms~\citep{johnen2018digital}\footnote{Firestorms are the sudden discharge of large quantities of messages containing negative word of mouth and complaint behaviour against a person, company, or group in social media networks'~\citep{pfeffer2014understanding}}. People may use the actions of others to help justify their own harassment behaviours~\citep{blackwell2018justified}.~\cite{phillips2015we} explored the relationship between trolling and mainstream culture, making the case that we not only have a trolling problem, but we also have a culture problem due to the way in which exploitation is viewed as a leisure activity within many societies; the internet has enabled trolls to organise, to mask their identities, and emotionally disassociate from the negative impact their behaviours have on those targeted.

Less organised forms of toxicity also exist online~\citep{wang2011regretted,warner2021,sleeper2013}.~\cite{wang2011regretted,sleeper2013} found people experiencing regret after posting comments containing foul or obscene language, negative and offensive comments, and messages containing direct attacks and criticism. Often these messages were posted during periods of anger and upset, suggesting a more hastily and less organised motivation for posting. This supports research that suggests, alongside the nature of the conversation, a negative mood can act as a primary trigger for trolling, and that ordinary people under certain circumstances can engage in this type of behaviour~\citep{cheng2017anyone}. 


\subsection{Toxicity interventions: from algorithm to interface}
Researchers in the areas of ML and NLP have developed algorithms for detecting different forms of abusive content such as toxic text (e.g.,~\cite{georgakopoulos2018}), hate speech (e.g.,~\cite{ketsbaia2020}), and cyberbullying (e.g.,~\cite{agrawal2018deep}). Context is an important consideration when evaluating message toxicity~\citep{blackwell2018}. To help improve the accuracy of detection systems, models have considered specific contextual factors including the nature of the broader conversation~\citep{pavlopoulos2020toxicity,menini2021abuse}), the sender and broad topic~\citep{ren2016context}, different dimensions of abuse (e.g., direct vs indirect)~\citep{waseem2017}, and sarcasm~\citep{davidov2010}. Yet, they are unable to understand complex contextual nuances within speech~\citep{caplan2018content} and are often used within industrial approaches to content moderation where consistency is prioritised over context~\citep{caplan2018content}. This can lead to miss-classification of content, and the embedding of structural inequalities into content moderation algorithms that impact Black people~\citep{haimson2021disproportionate,marshall2021algorithmic}, trans people~\citep{haimson2021disproportionate, dias2021fighting}, people with mental illness~\citep{feuston2020conformity}, sex workers, or content producers that relate to nudity~\citep{are2021shadowban,are2022autoethnography}, and others. Unsurprisingly perhaps, significant ethical concerns have been raised around algorithms used for moderation, including concerns of fairness, discrimination, and freedom of expression~\citep{kiritchenko2021confronting, vaccaro2020end}. 

\begin{figure} [t]
    \centering 
    \includegraphics[width=0.5\linewidth]{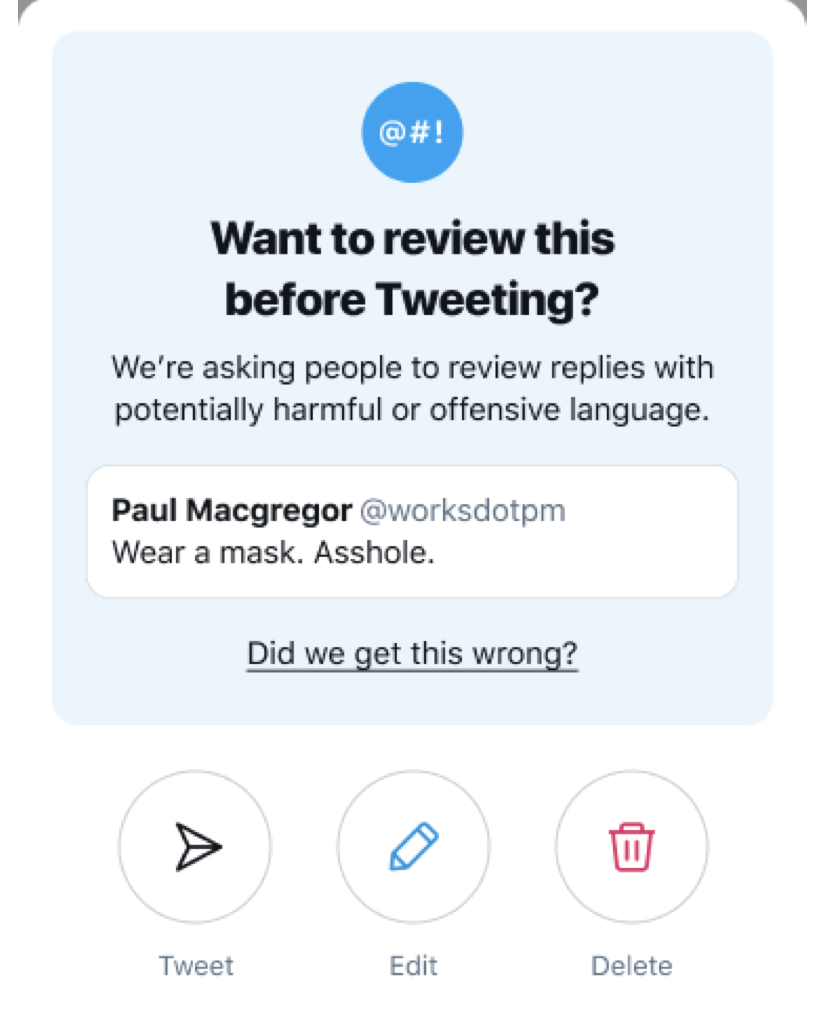}
    \caption{Twitter's proactive moderation prompt (cropped)~\citep{katsaros2021reconsidering}}
    \label{fig:twitter}
\end{figure}

\cite{kiritchenko2021confronting} highlight several alternative algorithmic approaches to reducing toxic speech, most of which shift how models are used within systems. These include using algorithmic moderation systems to nudge users into making better choices; importantly, this approach does not have to limit freedom of expression. Similar to this approach, several platforms have tested toxicity algorithms that use text-based nudges at the point of message sending, intending to reduce instances of abuse~\citep{katsaros2021reconsidering,techradar2021}. The dating app Tinder has implemented proactive moderation within private chats, nudging users when toxic language is detected, asking them:\textit{``Are you sure you want to send? Think twice - your match may find this language disrespectful.''}. Similarly, Twitter prompted users prior to Tweets being sent where \textit{``potentially harmful or offensive language''} was detected (Figure~\ref{fig:twitter}). 

An evaluation of Twitter's implementation of their proactive content moderation intervention suggests that, over a short period at least, it was effective at reducing offensive Tweets (9\% deleted, 22\% revised)~\citep{katsaros2021reconsidering}. The intervention had positive effects downstream, with fewer toxic Tweets resulting in fewer toxic replies; a finding reflected in other work~\citep{cheng2017anyone,chang2022thread}. When designing interfaces to support proactive moderation, researchers have drawn from the literature on reflective interfaces~\citep{royen2017,dinakar2012common,jones2012reflective}, and resonate with literature on reflective informatics~\citep{baumer2014reviewing,baumer2015}. For instance,~\cite{royen2017} explored different interface and interaction design opportunities to understand how they may reduce instances of online harassment by adolescents. They utilised different types of reflective text-based prompts (e.g., a prompt that relayed the potential harm of the message to the receiver). Across their conditions, they found a reduction in participants' intention to engage in online harassment. Similarly,~\cite{jones2012reflective} explored reflective interfaces in a fictitious social networking website, finding that they encouraged users towards positive behavioural norms. Others have explored browser plug-ins to prompt users when engaged in messaging on the Reddit platform, with the system providing users with guidance on the potential effect of their message on the civility of the conversation prior to the message being sent~\citep{chang2022thread}. 

\subsection{Manipulation and abuse of algorithmic systems}
Algorithms can be manipulated or abused by users. Some algorithmic manipulations are complex and require significant knowledge and skill to perform. For example, data poisoning involves attacker(s) contributing large amounts of data to a model to change the way it makes predictions (e.g., causing non-toxic speech to be classified as toxic). Other manipulations and abuse are less sophisticated. For example,~\cite{hastings2020lift} highlight the potential for students to manipulate AI team formation software into grouping them with their friends. Students learn the features the AI uses to cluster individuals into teams and complete the surveys to fit the model. Similar behaviours have been observed around NLP models. For example, manipulating a system that is designed to detect gender from the written text by perturbing input text~\citep{reddy2016obfuscating}, and perturbing toxic text prior to submitting it to a toxicity detection model to trick the system into reporting it as non-toxic~\citep{hosseini2017deceiving}. These input manipulation are often referred to as `adversarial examples'~\citep{szegedy2013intriguing}. Of course, manipulations like these are performed for differing reasons, sometimes with good intentions, and sometimes with bad -- points we will return to in our studies findings. In the case of toxicity manipulation, these are likely used to cause harm whilst evading automated moderation. 

In summary, research is being conducted to develop more accurate models to predict toxic speech. We also highlight research into how interfaces could be designed to prompt users into making better choices when communicating. However, there is a paucity of research on the process of integrating toxicity algorithms into socio-technical systems (RQ1), and the implications that this could have on user communication (RQ2) and potential risks they could expose users to (RQ3).

\section{Method}
\label{sec:method}

This section details the virtual design workshop method that we used to answer our research questions. It describes the design activities developed for probing and trigger discussion, rather than developing specific designs.


\subsection*{Participant recruitment}
Our virtual design workshops were conducted with three different stakeholder groups, and one user group. Below we describe the expertise of our three stakeholder groups. As these were small groups of individuals with specific expertise, we felt that reporting demographics was not appropriate and risked identifying individuals. We instead focus on collecting and reporting their expertise in an aggregated way which we describe below. For our user group (previous producers of toxic content), we did collect demographic information (Table~\ref{tab:demographics}) as, unlike our stakeholder group, we were not focusing on their expertise around content moderation. Moreover, while this group was small, they were taken from a large pool of potential participants meaning de-anonymisation was not feasible.

\textit{Group 1: Early career HCI researchers.} Using convenience sampling, we recruited five early career HCI researchers from within our university. We recruited this group as they were all experienced in participatory, co-creative, or user-centred design methodologies. Their insight into the methodology was a good testbed for our workshop activities and also provided general insight into how online moderation impacts users. All were living in the UK at the time of the study. Participants were offered either a voucher or charitable donation equivalent to GBP10/per hour (\~USD13). 

\textit{Group 2: Expert researchers and practitioners.} We used direct contact method to recruit both academic and non-academic experts. Experts were recruited to ensure we were building on current best-practice expertise that broadly addresses issues of online toxicity. Four experts were living in Europe at the time of the study. They had a combined experience of 30 years working or researching in the area of cyberbullying, online hate, censorship, and content moderation. Participants were offered either a voucher or charitable donation equivalent to GBP10/per hour (\~USD13). 

\textit{Group 3: Online content moderators.} We recruited online content moderators to focus the research on the practicalities of what content moderation algorithms must consider, and to learn from those that moderate content on a day-to-day basis across different types of online spaces. We recruited these participants via a number of online moderator discussion spaces. This was important to ensure that our research was implementable in settings where post-hoc moderation already occurs and that we embedded existing practical knowledge on content moderation into our work. The first author joined these groups and obtained permission from admins to advertise the study. A total of five online moderators we recruited, three were living in the USA and two in the UK at the time of the workshop. We collected details on their moderation experience. They had combined experience of 26 years moderating large online communities on Reddit, Discord, and Twitch. Communities that they moderated and were members of, related to: Science, Crafts and Hobbies, LGBTQ+, Gaming, Jokes, Relationships, Culture, and Anti-hate. Participants were offered either a voucher or charitable donation equivalent to GBP10/per hour (\~USD13). 

\textit{Group 4: Previous producers of toxic content.} We describe our user group as previous producers of toxic content. We engaged with this group as our ``end users'', i.e., those for which a proactive content moderation system could be used. They were important to include in the study since they are the main target audience of any intervention designed in the research. Obtaining their perspectives on what might change their behaviour is imperative if we want interventions to be implemented and useful to the target audience. The research team did not have direct connections to people previously engaged in sending toxic content, and so we used a recruitment platform (prolific.com). We deployed an initial screening survey to identify individuals who met our inclusion criteria of having previously sent harmful content that they later regretted sending. For this group, we collected basic demographic information through Prolific (Table~\ref{tab:demographics}). All participants were remunerated for their time via the Prolific platform (approximately GBP10/per hour (\~USD13).


\begin{table}[h!]
\centering
\begin{tabular}{|c|c|c|c|c|}
\hline
\textbf{} &\textbf{Age} & \textbf{Sex} & \textbf{Ethnicity simplified} & \textbf{Country of birth} \\ \hline
P1 & 46 & Male & White & Poland \\ \hline
P1 & 48 & Male & White & UK \\ \hline
P1 & 67 & Female & White & UK \\ \hline
P1 & 34 & Female & Asian & UK \\ \hline
\end{tabular}
\caption{User group demographic data}
\label{tab:demographics}
\end{table}

\begin{figure} [t]
    \centering 
    \includegraphics[width=1\linewidth]{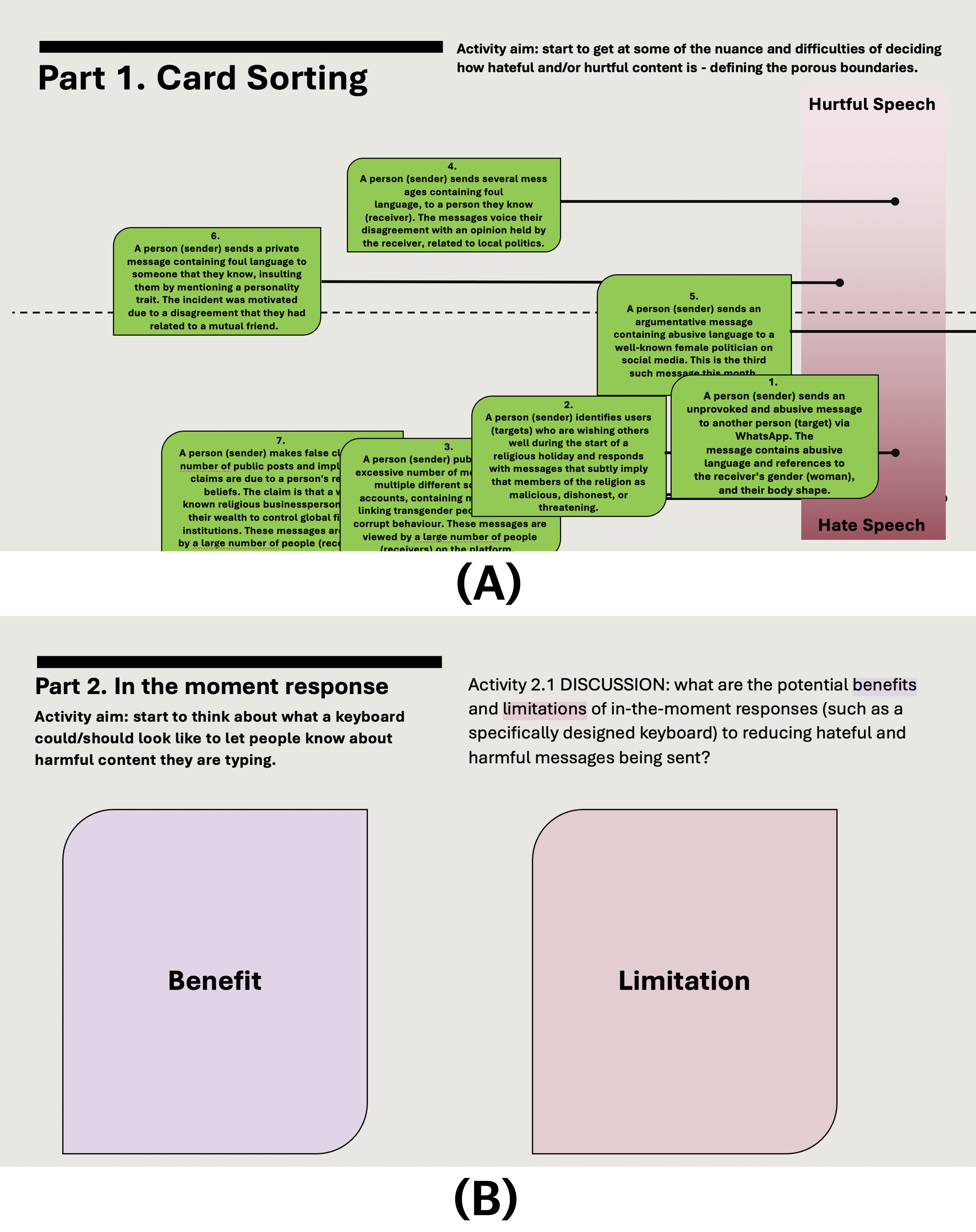}
    \caption{Part 1 and 2 workshop slides. Slide A supported a situation card sorting exercise; slide B supported an activity to elicit feedback on the benefits and limitations of proactive moderation systems built into a keyboard.}
    \label{fig1}
\end{figure}

\subsection*{Virtual design workshops}
In total, we conducted four virtual design workshops between 3 December 2021 and 18 January 2022, one with each of the four different groups. The workshops were facilitated by either the first or second author, with author four participating where possible. Each workshop included four to five participants and lasted approximately two hours. Workshops have previously been used in research to explore related topics (e.g.,~\cite{ashktorab2016} work on cyberbullying) and provide a means of engaging different stakeholder groups in discussions related to a shared goal~\citep{muller2012participatory}. They also allow for a variety of activities to be conducted with participants to facilitate a rich set of insights. We used this method to (1) explore and critique proactive content moderation interventions and (2) discuss design opportunities for potential future proactive content moderation interventions that responded to this initial critique. In this paper, we report on findings from the former. The workshops were supported by a slide deck that included information about the study, the workshop aims, and guiding questions, as well as interactive activities for participants (see Figure~\ref{fig1} for a preview of a number of the slides). During these activities, participants were invited to collaborate using a resource made available to them via a shared PowerPoint slide deck~\citep{lee2021show}, and to discuss their thoughts. 

\begin{figure} [t]
    \centering 
    \includegraphics[width=1\linewidth]{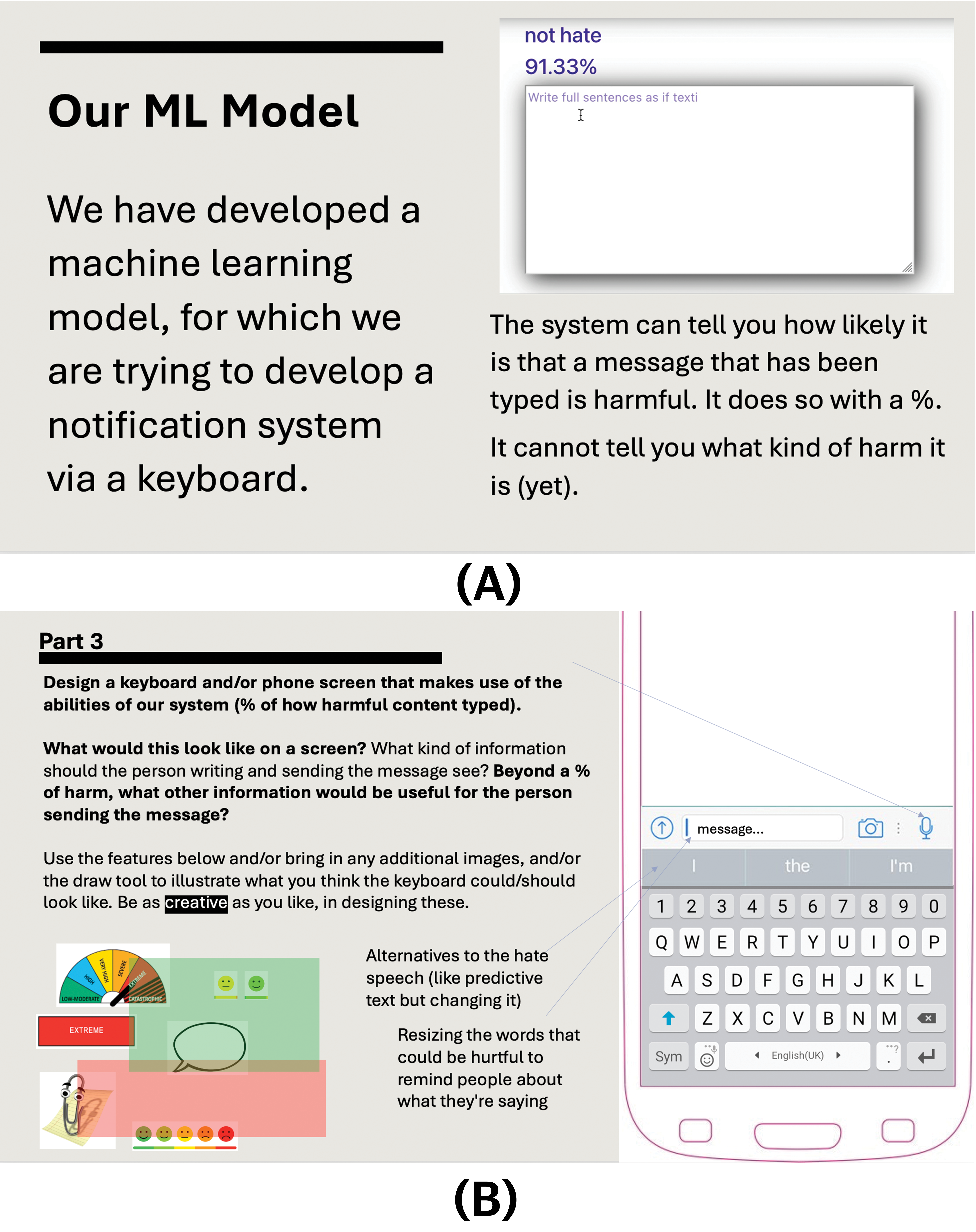}
    \caption{ML mode overview and part 3 workshop slides. Slide A provided an overview of the toxicity detection model; slide B supported participants in thinking about and discussing designs of a keyboard based proactive intervention that utilises the model.}
    \label{fig2}
\end{figure}

In the first part of the workshop, we used a situation card approach~\citep{muller2012participatory} and a scale of harm 
(see: Figure~\ref{fig1}A) to facilitate group discussion on the boundaries between different types of content (e.g., hateful and hurtful content). 
Each card contained a problematic situation containing toxic speech, and participants were asked to discuss where they felt the situation should sit on the scale. 
These cards were based on real-world examples that the research team collated. 
They helped us to answer RQ1 by drawing out the complexities of online harms related to online abuse and allowed us to explore the porous boundaries of language and legality. 
We included this task to stimulate thinking around nuances of problematic content during the latter design activities. Discussions build on activities and prompts such as these, embedded within the workshops. 
We then presented an algorithmic toxicity model to the participants and proposed the idea of integrating the model into a mobile keyboard to help reduce the production of toxic content. 
To help us answer RQ2 and RQ3, we prompted a discussion on the benefits and limitations of this approach (see: Figure~\ref{fig1}B) and recorded participant comments in two boxes so all participants could visualise and collaboratively add to them. 
The group discussions in this part of the workshop helped to lay the groundwork for nuanced discussion in the second part. 

In the second part of the workshop, we split participants into two breakout groups. There, we engaged them in a virtual collaborative prototyping activity~\citep{bodker2022activities} in which they were tasked with designing an imagined keyboard intervention to respond to toxic content production at the point of writing. To help us further answer our research questions within the context of the design of these systems, we asked participants to discuss their ideas, and to explain and reflect on their choices as they developed their ideas together. We provided them with a basic layout, and virtual drawing tools and ``props" that they could use to support their ideas (see: Figure~\ref{fig2}B). Designs were then discussed among the whole group, adding depth to the initial discussions in the breakout groups.

This study was approved by the ethics boards of both Northumbria University, and the National Research Centre on Privacy, Harm Reduction and Adversarial Influence Online (REPHRAIN). The main considerations made relate to the potential for participants to discuss issues that might be considered hateful or harmful, and for this to have a negative effect on participants. To mitigate this, we instructed participants not to discuss personal experiences of abuse. At the end of each workshop session, all participants were sent links to a number of well-respected charities that support people who experience instances of bullying and abuse.

\subsection*{Data analysis}
Each of the workshops produced three transcripts (one for the main discussion, and one for each of the two breakout groups in the second part of the workshop). We used a collaborative constructionist thematic analysis on all transcripts, following~\cite{braun2006using}'s six phases. Initially, author two inductively coded the first set of transcripts to develop an initial set of codes that were visually represented on a shared online whiteboard. Following this, authors one, two, and three deductively coded the remaining transcripts using the shared whiteboard. This allowed for continuous collaborative discussion on how transcript data was being coded. Using a shared online whiteboard, all authors asynchronously developed and reviewed themes. Adding to this asynchronous analysis, all authors discussed theme development and used the online whiteboard to support them in defining and naming themes before collectively producing a report. Drawing on~\cite{walny2011follow}'s learning about sketching in software development and~\cite{sturdee2019sketching}'s sketching as future inquiry, we augmented our thematic analysis with the sketching of design ideas produced by participants. This allowed us to visualise potential futures, adding depth and insight into our thematic analysis. Our thematic analysis was reflexive and understands that we have constructed our below arguments from the data. As such, we do not present quantitative information on our codes or themes.

\section{Findings}
\label{sec:findings}

The findings that we present in this paper relate to three themes that were developed as part of our analysis that focused on the critique of these systems. These themes are summarised along with sub-themes in Appendix Table A\ref{tab:summary}. The three main themes are: (1)  the different contextual factors (which explores different contextual elements that relate to how toxicity is understood), and how they can help us design better algorithmic socio-technical systems; (2) the different end-user groups involved in the creation of toxic content, and (3) the ways such algorithmic systems could be abused or circumvented by different user groups. Where we refer to participants, we use `R' to refer to HCI researchers, `E' for experts, `M' for moderators, and `P' to refer to previous producers of toxic content.

\subsection{Contextual factors impacting on perceptions of toxicity and effectiveness of the system}
Participants in all groups discussed how different contextual factors might affect how proactive content moderation systems work, their effectiveness, and in turn how this would impact on users. During our analysis, we discussed various aspects related to the severity of toxicity, how toxicity is understood, and how it should be proactively moderated, which relate to different contextual factors. Below, we present four factors our participants discussed including interpersonal contextual factors, scale of abuse, platform culture and affordances, and social histories of oppression and power structures.

\subsubsection{Conversational and interpersonal histories} Conversations containing toxic language are not sent in vacuums, they have histories - both interpersonally and publicly. While certain messages in and of themselves might be classified as toxic by an algorithmic model, others may only be viewed as toxic when considered alongside the broader context of the conversation in which it occurred. P1 highlighted this, saying: ``I think if there's a course of behaviour rather than a single act, that suggests [..] a heightened hate element to it''. 

We also found a difference between conversations and the language used between friends, people within a cultural or social in-group (e.g. LGBT+ fora), a workplace, or cultural community; there were concerns related to how algorithmic models would account for these differences. P1 said: ``Boundaries will be different in different cultures and so how do you factor in all those different variables to make something that actually does the job for everybody?''. 

E1 made a distinction between friends and strangers highlighting how in-group conversations may contain language that a model might inappropriately consider to be toxic due to the close relationship shared between friends, saying: ``the impact is not only dependent on the content of the message. If I say `bitch' to my friends, he or she will not be offended by that. But if I say `bitch' to somebody I don’t know I think the impact will be much larger''.

Similarly, content moderators discussed the need for freedom to share toxic content to build in-group solidarity through humour, for example in an LGBT+ forum. Yet, they also discussed negative outcomes when applying the same moderation rules to extremist groups (e.g., white supremacists). 

While individual conversations and relationships have histories, so too do public debates. One participant pointed to an example of where a public debate contained a history that could result in heightened emotional responses from people involved in related discussions. Here, content could be flagged as toxic without considering the nuances of the public debate and its threat to individuals involved. E4 said: ``a lot of female politicians are very openly against sex work in a way that might bring in policies that actually make sex workers' lives harder and therefore a very toxic kind of debate cycle ensues, so I can see people being scared about these policies coming forward and being very argumentative and almost insulting''. 

\subsubsection{Scale of abuse} The number of individuals involved in sending toxic content to or about an individual can impact greatly on the harm caused. During part 1 of the workshop, M4 spoke of the need to consider the scale of the abuse, and not just the toxicity of individual messages. They said: ``even if the single messages aren't hateful once you get enough people sending someone messages, the impact gets a lot greater''. M1 also commented on this in relation to misogynistic behaviours, saying: ``when there's 100s of people sending 3 messages a month to the same person, that volume, it makes it much more explicitly, uh, misogynistic, hateful''. Where moderation is performed proactively based on the content being typed, there were concerns around the algorithms ability to consider prior messages within the broader context of the behaviour. Aligned with~\cite{marwick2021morally}'s notion of morally motivated network harassment, messages in and of themselves might not be classified as toxic by an algorithmic model, but may be part of a network-level harassment campaign, or written in an attempt to recruit others into engaging in network-based harassment. When considering one of the cards in part 1, P1 said that this would increase the toxicity of the message, saying: ``I think that's far more painful because you're not only carrying the hate yourself, you're trying to recruit others into your hatred as well''. Further to this recruitment into their toxic way of thinking, another participant explained how such actions can lead to attracting further abuse, turning into a targeted campaign that can snowball out of proportion. E4 said: ``because this stuff is also then picked up by certain media outlets [in a] kind of regurgitation [of] the message, [...] essentially just making online abuse news.''.

Many of these discussions were entwined with discussions related to the size of the audience and how audience size could impact the harm that resulted from the content. The audience size was discussed in relation to the effect of the speech, considering the longer-term implications of the speech (e.g., how it might change views, opinion, beliefs over time), as opposed to short-term impacts. M3 said: ``sending the abusive message to another person is, that's a dialog between you [..] I'm not just looking for the people that are trans or Jewish, or is Islamic. If they're being Islamophobic, they're looking for those neutral people, they're looking for those people who are there sitting on the fence, which you know, are ready to fall one way or the other''.

Further to the size of an audience, participants within the Expert, Producer, and Moderator groups pointed to a difference between potentially toxic content that was available publicly and that which is visible only by individuals (or even one individual) in a private setting (M3). P4 linked this to the reaction that a message might cause: ``If you were to send this [toxic message] only to this individual, then it's sort of personal. But if it's on a wider forum [...] then maybe you are starting towards hate''. Finally, creating content in a public space means that the person is ``reiterating the message over and over'' (P3), increasing its potential impact.

\subsubsection{Platform culture and affordances} Participants in the Producer and Moderator group discussed a diversity of platforms such as Whatsapp, Twitter, Facebook, and Reddit. They pointed to these as having different dynamics, not least because the purpose and user-base differ across platforms. Participants pointed specifically to a difference in political conversations across different platforms, highlighting the perceived polarised nature of some environments. When discussing one of the cards in part 1, P1 said: ``If you're on social media, things like politics have become incredibly tribal [...], on Twitter it is very, very difficult to just have a level headed political conversation of any depth with anybody''.

Many of the discussions were entwined with the affordances of platforms. For example, Twitter's trending topics feature and recommendation algorithms would make it easier for transient trending topics to be hijacked and used to spread more subtle forms of hate. In part 1 of the workshop, M4 described this in relation to a Twitter trend involving a comedian perpetuating transphobia. This resulted in people using pictures of the comedian to target LGBTQ users; they said: ``it's almost impossible to be ahead of those trends [...] unless you're like constantly looking at whatever is trending on Twitter''. 

\subsubsection{Social histories of oppression and power structures} Conversations and content shared online are part of our lived experience, both in our offline and online developments. In turn, this means they do not exist in socio-political or power vacuums. What~\cite{hooks2003teaching} describes as ``the capitalist imperialist white supremacist patriarchy'' continues to shape our post-digital world - including the online content we share and how algorithms are trained to understand them. This means, as one of our participants in the Moderator and Expert groups points out, that flagging sentences like ``men are trash'' and ``women are trash'' in the same way does not acknowledge the relation of these sentences to ``centuries of misogyny'' (E4). Even if a message is not outwardly visibly harmful to some, M2 said that: ``that message may still be hateful. That message may still be targeted or rooted in misogyny and in being hateful to women''. While our examples relate to misogyny in particular, these kinds of power structures relate to all axes of oppression such as race, sexuality, and disability. 

\subsubsection{Summary} Addressing RQ1 related to the broader challenges of embedding toxicity-detection algorithms into socio-technical systems, we find participants across all groups expressed that simply using the textual and conversational context of toxic messages in algorithms is not enough. When we start to use toxicity algorithms in socio-technical systems such as proactive moderation systems, additional contextual factors shape what both interface and algorithm designers must consider. We do not believe that a single design can address each of these factors, but we should be conscious of them as and when systems are developed. We are far from the first to point out contextual factors of content moderation~\citep{caplan2018content, gillespie2018custodians,scheuerman2021framework}, or the first to show that frequency of messages are an important component of severity of online hate~\citep{Jhaver2018,marwick2021morally}, or that in-the-world power structures shape online harms~\citep{schoenebeck2021drawing}. However, we do identify the presence of these contextual factors within a proactive moderation paradigm, and aggregate and arrange these factors to help add a new kind of interest in this area.

\subsection{An end-user continuum of intention to send toxic messages}
Here, we present our analysis of how participants across all groups expressed who may be using proactive toxicity-reducing interventions, presented across an end-user continuum from those unaware they are harming others through their messages, to those with high levels of intention to commit acts of harm. We start from users with low intention to cause harm and increase the intentionality across the five groups. We hope that by presenting these various user groups, future interventions can become more specific to use-cases and users.

\subsubsection{The unaware and open to learning} On the lower end of the continuum, participants in the Moderator, Producer, and Expert groups talked about people who may be unaware of the language they are using when communicating. For instance, people ``accidentally using dog whistles [they] genuinely don’t know about'' (M3). Participants discussed the potential for a proactive moderation system to be effective in highlighting harm to those unaware that they are producing it, and providing a mechanism to support them in learning. For example, P4 said: ``Some people might not have the intuition to just realise it straight away because probably if they had they would have not written the message in the first place''. 

Especially for those unaware of the reason for their messages being considered hateful, participants highlight the importance of the intervention to act as a form of support, offering not just prompts, but information to educate those to help them avoid engaging in this form of toxic speech production in the future. However, there was a recognition that this was not easy for people. P3 spoke of their own journey, saying: ``that's something that I've had to learn, and that's something that takes real dedication and mindset change for an individual, and they have to want to better themselves as a person and develop themselves as a person''.

\subsubsection{The emotionally triggered} There was also a recognition amongst participants in the Producer, Researcher, and Expert groups that people who do not regularly engage in toxic content production may send a toxic message after being emotionally triggered. P3 spoke of their own experience, saying: ``if I was to say something hurtful to someone. It's very likely that it's almost like a rebound defence mechanism: ‘Somebody said something to me and therefore I want to say something back’. It's that kind of mentality''. In these moments when people are emotionally triggered, they may be more susceptible to engaging in toxic speech production. There was a concern amongst participants that this behaviour may become a source of regret, highlighting the potential harm that toxic speech can have for both those that are targets of the speech, and those that produce it. 

\subsubsection{Those using toxicity as a form of emotional arousal} As we move towards the higher end of the continuum, participants in the Expert group discussed people who engage in problematic and toxic interactions for emotional arousal (e.g., to have fun), a finding that is reflected in prior work~\citep{navarro2021trolls, cook2018under, buckels2014}. Highlighting this, E1 said: ``apart from anger [..] we also see that boredom is often a trigger and then being aggressive or joining in bullying or cyberbullying is actually just a way of having fun. Or yeah, being stimulated or being excited''.

\subsubsection{Those playing to an audience} People may also act or play up to an audience~\citep{golf2022feeding}, engaging in toxic speech production as a means of showing off. Both the Expert and Moderator groups discussed this intention. M4 talked about this explicitly, saying how: ``once you add an audience, I think bad actors do weirder shit''. There were concerns that a proactive moderation system such as that built into a keyboard may make this preformative behaviour worse, allowing those involved to highlight how harmful their communications are being in a more quantifiable way, which we come back to later in the paper. 

\subsubsection{The determined and organised} Towards the high intention end of the continuum, participants in the Expert, Researcher, and Moderator groups talked about people who may change their behaviour but not their views and beliefs. For instance, R5 said: ``you might stop people from [...] sending these things, but it might not make people believe differently''. There was a general view that it would be more difficult to change the mindset of more stubborn actors, especially: ``political activists or [those] who are part of a certain community and do this in an organised way'' (E1). Similarly, participants were concerned about the effectiveness of any intervention aimed at people involved in organised toxicity campaigns. E4 said: ``For people who actively set out to cause harm. [...] if you're going there on your keyboard to just be hurtful or hateful or harmful getting that message [...] if anything, it might just spur you to do more''. 

Participants highlighted how those that are determined and organise would be unlikely to consider using interventions at all, and if they did were unlikely to change their behaviour, or may even use it to further their abuse - an issue we come back to in the next section. 

\subsubsection{Summary} We know from prior work that both organised and coordinated groups~\citep{mittos2020a,mittos2020b,mariconti2019, zannettou2018gab, chatzakou2017hate, chatzakou2017}, as well as everyday people~\citep{cheng2017anyone} engage in uncivil discourse, but we lack a more nuanced understanding of different users that may engage in this form of behaviour, and how these groups may respond to proactive forms of moderation that promote user-agency. Addressing RQ2, our findings present as a continuum of potential users of a proactive moderation system that presents information from a toxicity algorithm. We feel it is important to state that people are likely to move between various different aspects of the continuum, based on their purpose for engaging with others online - for example, a person may simultaneously want to learn about how to change their behaviour, while also having instances where they are emotionally triggered. We come back to this point in the implications for design. Throughout the continuum, there are several points where participants point towards the potential for abuse of such systems, which we expand on below. 

\subsection{Abuse and manipulation of applications which embed toxicity models}
Despite the presumed positive impact of a keyboard that highlights toxicity levels of a message, such systems are rife with potential for abuse - and may have the opposite effect to that which was originally intended, for example, by providing a tool to help validate attacks, promote the gamification of toxicity, allow people to circumvent detection or manipulate the system. 

All groups talked about the potential for abuse of an intervention that presents algorithmically-determined scores or information about a  degree of toxicity, similarly~\cite{jhaver2019human} and~\cite{chancellor2016thyghgapp} have explored various manipulation and gamification practices among AI-based moderation tools. The ``abusability'’ our participants discussed expand these literatures as they link to the above end-user continuum, as it does not refer to all of these potential users equally. Similarly, it relates to the contextual factors as these may shape how people use the system. Participants across all groups highlighted features of the interface which could lead to reverse effects, in particular: ``making sure you're definitely not showing the percentages to gamify it'' (M3). One design choice discussed was ``not giving you value judgements on every message you ever send. It's just flagging problems to you when they come up'' (M4). Ultimately though, this is not only an interface issue - our participants discussed repeatedly in all the workshops that the abusability of such systems directly relates to the toxicity detection algorithms themselves which determine the kinds of features that can be implemented via design interventions, such as interfaces. Below, we outline four ways in which these systems could be abused: through validation, gamification, model manipulation, and circumvention. 

\subsubsection{Validation}
Participants in the Moderator and Producer groups pointed to how an intervention that provides toxicity feedback could be used to validate attacks. M2, for example, used the insight about toxic content they have moderated to argue that people who want to cause harm online may interpret toxicity scores with the inverse reaction as intended. They said: ``The value of seeing that you know you're 80\% likely to be saying something hateful is actually, may even have a reverse psychological effect because I got validation, I'm being hateful, the machine knows I'm being hateful''. Another moderator addressed how the design may facilitate an exploration of future hurtful content: ``now you've drawn attention to, possibly, a stereotype or insult'' (M3).

\subsubsection{Gamification}
The embedding of toxicity algorithms into a proactive moderation system could also lead to a reverse effect through gamification, as discussed across all groups. For example, M4 pointed out: ``People would take the most mundane things and turn [them] into [a] game''. They also pointed out that people who want to cause harm may interpret a toxicity score (whether presented in percentages, colour-coding, or any other form of feedback) as a competition: ``Who can be the most hateful with a percentage score?'' (M4); or as another moderator said to share their progress ``screencap it and just say look how, how hateful I am'' (M1). Again, drawing on the experience of moderating online forums, M4 even pointed out how people who want to cause harm may even ``rack up'' their ``negative internet points'' or as P2 adds: ``Some people would actually relish having that needle go up to the top. [...] They're looking for that. Uh, almost confirms to them that they're accomplishing, what they want to do inside themselves''.

\subsubsection{Model manipulation}
Participants in the Moderator group spoke about ways in which they themselves and their communities could help to improve a toxicity detection model, making it more accurate by better understanding language and norms within their communities. M3 who had some high-level understanding of how machine learning models worked, talked about allowing communities to help train a model by highlighting words and sentences that the model should not flag as being hate. She said: ``I think if at [..] the subreddit level of r/[anonymised] [I] was able to see a list of all the stuff that gets flagged and we could say ``like, yeah, that's not actually hate speech that's not hate speech, that's not hate speech". I think that would probably be useful feedback for like the whole model''. 

However, as part of the same discussion M3 and others highlighted the potential downsides to this type of approach due to the potential for this to become an avenue for abuse. Again, M3 said: ``You would also open it up to bad actors intentionally providing it bad data [..] and communities will just go through and flag everything they don't agree as not hateful''. 

\subsubsection{Circumvention}
Furthermore, participants across all groups pointed to the potential users on the ‘determined and organised’ side of the end-user continuum, and how they are likely to find ways of circumventing the system. P4 puts this matter-of-factly: ``Obviously, apart from the obvious that people who want to cause harm and hurt, they would just circumnavigate it or or they would not have this keyboard''. The experts focused more on the impacts of toxicity detection algorithms if they were implemented in a way that people were forced to use them, E2 said: ``How far can I push the system? What kind of combinations of things can I say that maybe go undetected? Or how can I just? I don't know find another way to say something hurtful in a way that just the system doesn't recognise it''. 

E5 expands on the above, adding that this behaviour may only arise after a user becomes frustrated with the platform (which points to a broader group of people on the end-user continuum than just those who are determined): ``if they start to get frustrated at the fact you know you keep warning them and highlighting things in their message, they might then go to just send a voice note instead''.

\subsubsection{Summary} Helping us to answer RQ3, we highlight several ways a toxicity detection algorithm (embedded into a proactive moderation system) could be abused by end-users and have the reverse effect for which it was designed. As with any intervention, such potential negative outcomes need to be considered and (ideally) mitigated in design; and measures should be put in place to allow for changes post-deployment based on user feedback.

\section{Discussion}
\label{sec:discussion}

Above, we have presented our findings systematically in a way that researchers and those who design and implement proactive toxicity moderation tools can make use of them. We presented context-factors (RQ1), the continuum of end-users (RQ2), and ways in which tools may be abused and/or manipulated (RQ3). We argue that with this collected knowledge, we will be able to design more nuanced, context- and community-specific interventions that take into consideration at least some aspects of abusability; to work with a harm reduction model. Below, we present three points of discussion before presenting considerations for designing proactive moderation systems that embed toxicity algorithms. 


\subsection{Algorithmic inequality and the importance of context}
We know from prior work how inequitable many of the existing social media content moderation processes are~\citep{haimson2021disproportionate,marshall2021algorithmic}. This inequality results in certain groups experiencing higher instances of content removal and account restriction~\citep{haimson2021disproportionate}. Applying moderation at the point of sending can help to alleviate some of these issues, as it affords users with increased transparency around what content might be considered toxic or inappropriate prior to sending~\citep{kiritchenko2021confronting}. This allows users to curate their content to reduce the risk of it being subjected to unjust content moderation practices. In exploring aspects of context around these systems, many of the concerns related to context were woven into concerns of inequality. Equality and representation in content moderation systems have been explored by~\cite{vaccaro2021contestability} who identified the need for systems to account for local differences or norms and behaviours when evaluating text toxicity, supporting our findings related to contextual consideration around interpersonal relationships such as in-group conversations. Whilst proactive moderation helps to address issues of representation and contestability~\citep{vaccaro2021contestability} by making moderation more transparent by being embedded within the user interaction, participants rarely considered this, and instead focused on the injustice felt if the system was to prompt them incorrectly, treating them (and their communities) unfairly. While these systems may not directly curtail freedoms of expression by restricting what a person can, or cannot say, there were concerns of~\textit{algorithmic conformity}~\citep{liel2020if} occurring that could indirectly curtail these freedoms over time~\citep{feuston2020conformity}. This involves individuals and groups adjusting their behaviours to conform to what the algorithm considers to be~\textit{good} and that this could be made worse through abusive model manipulation. Highlighting the duality of these systems,~\textit{algorithmic conformity} is a desirable outcome of the system for those committing acts of harm by producing toxic speech. Yet,  ~\textit{algorithmic conformity} is wholly undesirable for those marginalised groups whose speech is being inappropriately flagged as toxic as a result of the algorithm being unable to consider the many contextual complexities that we highlight as part of our research.


\subsection{Negotiate public good with personal regret}

Integrating a toxicity algorithm into socio-technical proactive moderation systems has some individual benefits. For instance, it allows users with low intention to commit acts of harm (e.g., the unaware) to become aware of the unintentional harm embedded within their messages. This would allow them to make a more informed choice and would help to reduce instances of regretful messaging~\citep{warner2021, sleeper2013,wang2011regretted}. Moreover, it re-enforces intentionality within the message and limits a person's ability to claim ignorance. Yet, this same mechanism also has the potential to exacerbate harm where both the sender and receiver are aware that messages are being moderated prior to being sent. Here, the system is likely to generate what~\cite{donath2007signals} refers to as a `signal', signalling a higher intention to cause harm where the sender has been notified of the toxicity in their messages and sends it regardless. Unintentional adverse effects from safety nudges have been observed in prior research with teens~\citep{obajemu2024towards}. For example, content warning labels increasing a teens ``paying more attention to it [the censored content] than if not censored''. 


Our work also highlights user groups (e.g., determined and organised users) for which this moderation approach is unlikely to have any direct effect, but may have indirect downstream effects such as changes in community-level norms~\citep{seering2017shaping, cheng2017anyone}. We have seen some early evidence of this from Twitter's embedded intervention where they identified fewer toxic tweets being created as a result of their intervention prompt (direct effect), as well as fewer replies to Tweets that received a prompt (indirect effect)~\citep{katsaros2021reconsidering}. Similarly,~\cite{cheng2017anyone} found exposure to trolling behaviours online can act as a primary trolling trigger. Yet we do not know how platform norms may reshape as a result of these interventions, and whether the knowledge by both the sender and receiver that these interventions are in place, will result in broader changes in behaviour. Of course, there remains some ambiguity between the sender and receiver as to whether the sender's message met the algorithmic threshold to cause a toxicity prompt. However, we know from prior work how people develop mental models (or `folk theories') around the functionality and capabilities of algorithms in online social environments~\citep{devito2017algorithms,devito2018people} and change their behaviour as a result~\citep{karizat2021algorithmic}. People may therefore start to loosely predict (or test) when a message has met the threshold for being considered toxic. If the intervention is embedded into the platform itself as opposed to being a stand-alone keyboard (e.g.,~\cite{katsaros2021reconsidering,techradar2021}), a flag that a message is toxic could be embedded within the sent message. Drawing on~\cite{erickson2000social} concept of socially translucent systems, this would enhance visibility, awareness, and accountability within the system by reducing ambiguity within the communication to foster community-level norm change and social learning. 
However, we must also be mindful of the potential for this type of system to act as a signal that could be exploited to exacerbate harm. 

 
\subsection{Circumvention, manipulation, and abuse of proactive moderation systems}
Drawing on our findings and those from prior research, it is clear there is potential for proactive moderation systems to act as interventions to support people's self-presentation and reduce toxicity. For example, helping to reduce the risk of people posting and sending something that they later regret that might have otherwise harmed their online identity~\citep{sleeper2013,wang2011regretted,warner2021}, and harmed receivers of that content. However, our work also highlights a number of potential abuse and manipulations of these systems which would be critical to consider when thinking about their design. Supporting prior work in HCI~\citep{chancellor2016thyghgapp, feuston2020conformity, gerrard2018beyond,jhaver2019human} we find concerns related to the circumvention of these moderation systems, through the purposeful manipulation of content. NLP researchers refer to these manipulative inputs as~\textit{adversarial examples}. They typically perturb the input into the model with the intention of causing the model to make an incorrect classification~\citep{liang2017deep,jia2017adversarial,szegedy2013intriguing, hosseini2017deceiving}. Our participants also raised concerns related to model manipulation where users are able to provide feedback to a model in a way that manipulates how it behaves. Prior research has explored this form of model manipulation around Twitter's `trending' algorithm showing how compromised and fake accounts could work to manipulate this platform feature~\citep{zhang2016twitter}. Other forms of manipulation have also been observed in relation to user content `flagging' where users are able to flag a piece of content (e.g., flag a post as abusive). ~\cite{thomas2021sok} refer to the manipulation of these features as `falsified abuse flagging' and can result in a moderation system incorrectly removing content that it considers abusive as a result of false flags that have been sent. ~\cite{fiore2012user} uses more militaristic language to describe group actions to remove content, describing these behaviours as `user-generated warfare'. These behaviours have been observed in the world~\citep{matias2015reporting} with activists using flagging functions to close porn performers' Instagram accounts~\citep{clark-flory2019}, and provide political groups with a means to remove their rival's Facebook content and restrict access to their account~\citep{yaron2012}. These highlight the extent to which certain groups (or `end-users') will go to misuse systems for their own harmful intentions. 

Extending this prior work, we identify further ways in which proactive moderation systems could be abused, and in doing so we highlight their potential to exacerbate harms. Like many technologies that are designed with good intentions, there often exists the possibility for technologies to be misused to cause harm.~\cite{hovy2016social} in their position paper on the social impact of NLP systems highlight the need to be mindful of `dual use' of systems and the negative social impact NLP systems could have through their use. Distinct from prior literature, we offer new insights into how proactive moderation systems could be misused to validate hateful comments and to facilitate harmful forms of gamification. Within an educational context,~\cite{andrade2016bright} highlight the risk of undesired competition as users feel forced to engage in competition with peers. Similar concerns have been highlighted by developers and designers of social VR spaces, concerned that gamified or incentive systems may be ``corrosive to community" by creating a culture of competition and envy~\citep{mcveigh2019shaping}. Our work resonates with these concerns, whilst also pointing to a form of inversion of the intended goal of proactive moderation systems. We highlight the potential for certain end-users to appropriate these systems to amplify and validate their abuse through gamification. 

\subsection{Considerations for designing proactive moderation systems that embed toxicity algorithms}\label{discussion:consideration}
Many of the issues we outline above are not only HCI, Interaction Design, ML, or NLP issues. Instead, they point to the deeply interdisciplinary nature of proactive content moderation. We argue for movements towards better systems, we need more proactive and deep collaboration between HCI and ML researchers and developers. Below, we present three considerations for working in HCI-ML collaborations. We understand that some of the considerations are more design oriented while others are more targeted towards ML or NLP developers. This was a conscious choice, since the work of these two end-users are deeply intertwined (both in academia and industry), and increasingly must collaborate. Ultimately, we argue for increased and respectful collaboration between the two disciplines.

\subsubsection{Presenting the output of the model}
Our findings suggest the need to consider how the output of the deployed ML model is being integrated into the front-end design of the system. Misuse of a system through gamification could be reduced where the output of the ML model is abstracted into a single dimension (e.g., toxic or not toxic) as opposed to providing a more granular output (e.g., 70\% toxic). Moreover, our work also suggests the need to consider how the output of the model is described to help reduce the risk of the system being used to validate hateful attacks. One approach could be to introduce an element of ambiguity of information into the system~\citep{gaver2003ambiguity}. For instance, rather than describe the output as being `toxic', describe it as being `inappropriate'. It is important to highlight here the trade-off between transparency and reducing the risk of exasperating harm. While there is a push towards increasing transparency and explainability of AI systems more generally (e.g.,~\cite{binns2018s}) as well as in AI systems that support moderation~\citep{suzor2019we}, we should also consider how increased transparency impacts risks of misuse. For example, increasing transparency within proactive moderation systems that rely on the same algorithm used for post-publication moderation may provide users with a tool to curate and evaluate the toxicity of messages prior to sending, to avoid automated flagging of content by post-publication moderation systems.

\subsubsection{Building feedback into the system}
Prior work suggests that where users are able to provide feedback on the decision of an automated moderation system, this can increase trust through an increased sense of user-agency~\citep{molina2022ai}. Our findings resonate with this, with our participants (and their communities) wanting some involvement in the process of training the model, and in improving the accuracy of the underlying toxicity detection model. This approach could help to address some of the concerns related to contextual understanding, and local language norms. However, we identified a concern that any such system could itself be misused as has been previously discussed in relation to `user-generated warfare'~\citep{thomas2021sok,clark-flory2019,are2022autoethnography,fiore2012user}. This could result in an increase in false-positives related to posts of already marginalised users~\citep{bender2021dangers,noble2018algorithms}. One way of mitigating this type of misuse would be to deploy individual or community-level models that are refined for individuals or communities. This would allow the user to flag false-positives to help the accuracy of the model in a way that is bespoke to them, as opposed to one that impacts broader model predictions across an entire platform. Yet, this type of approach may act as a double-edged sword. While this type of model customisation (either at the individual or community level) may be `good' for some communities (e.g., preventing posts discussing abusive behaviour in an abuse support fora being flagged as toxic), as discussed by our participants it could be abused to exacerbate harms that manifest in problematic communities (e.g., to legitimatise racism in online fringe communities).

\subsubsection{Interdisciplinary working from model to implementation}
Drawing on the literature we outline earlier in the paper, we understand that ML and NLP experts’ assumptions for better hate-detection and reduction systems often relate to having enough data, the right kind of data, or even the right labelling of this data~\citep{pavlopoulos2020toxicity,waseem2017}. Counter to this, hate-content reduction systems that draw primarily on design and HCI literatures rely on the assumption that what is required is the ‘right’ interface, data-representation, or behaviour change system to do this~\citep{jones2012reflective, royen2017}. It should be without saying, that both of these issues should be addressed when implementing new models as part of toxicity-detection algorithms in proactive content moderation systems: they are both a ML and/or NLP and an HCI issue. As such, we argue that it is important to also reflect on the design process of creating interactive systems that turn toxicity algorithms into socio-technical systems. 
When bringing these two audiences and problems into conversation with one another, and presenting this as a potential system to a variety of potential users of such systems as we have done, we learn that this becomes not only a socio-technical but also a socio-political issue; where hate-detection and reduction systems are embedded in messy contexts (as we outline in our finding on contextual factor) which relate to people in different ways (as we expand on in the end-user continuum), and one which lays bare multiple misuses of the system which can lead to increased harm in some instances (as we lay out in our section on the abusability of such systems). 

Considering the contextual complexities highlighted in this work, it is unlikely that any one discipline can resolve the socio-political and socio-technical issues highlighted. Therefore, we argue for the need for ML and NLP researchers and Design and HCI researchers to genuinely work together to create systems that are greater than the sum of their parts. When designing any socio-technical system that utilises an algorithm, careful consideration should be made towards how multidisciplinary teams work together to develop solutions. 
For instance, care should be taken to avoid adding unnecessary constraints to design and development processes. As an example, some machine learning and NLP models are developed before the problem space and user needs are fully understood. E.g., where ML or NLP experts first develop an algorithm that is then later implemented into systems by designers and software developers. This model of development (algorithm first, design second; or the use of resources such as perspective API to pre-build systems) can constrain front-end developers and interaction designers by forcing them to develop user interfaces and experiences that conform to both the input and output requirements of the developed algorithmic model. We do not say here that designers are always or entirely constrained by using only existing models. Wizard of Oz testing, Design Fictions, and other design methods allow us to explore potential new design-led avenues for proactive content moderation systems and toxicity detection algorithms. However, when implementing systems in-the-world, we rarely see processes that are truly iterative at both the ML and Design stages. As such, we argue for more genuine interdisciplinary (or at least iterative multidisciplinary) working: where HCI and ML/NLP experts work together prior to a model being developed, so considerations can be made as to how models are trained, the type of data used to train them, and approaches for labelling of data that better reflect the end-users that the system is designed to support.

\section{Limitations and future work}
While we believe our findings are useful for researchers working across and at intersections of HCI and Data Science, we also understand that the findings have limitations. Firstly, our sample size was small and while this was sufficient for this type of qualitative research in allowing us to develop these insights~\citep{caine2016local}, we are unable to generalise our findings. Moreover, we did not collect certain demographic information across our non-end-user stakeholder groups and so we are unable to comment on findings related to, for example, gender or ethnicity. All of our participants were over the age of 18, and so we are unable to learn how these types of systems might be viewed by younger adults; however prior work has explored different a set of safety nudges with teens~\citep{obajemu2024towards}. What we present within this research is an analysis of the experiences and expertise that our participants have shared with us, we do not present specific design choices nor specific designs that participants created during the workshops. While there are concerns about harmful misuse of proactive moderation systems through gamification and validation of hate, it would be important to understand the extent this might occur. As part of a broader project, we have developed a test social media and instant messaging environment to help us understand how users respond to moderation systems that provide granular model feedback (e.g., ``Your message is 70\% toxic"), vs more abstracted feedback (e.g., ``Your message is toxic"); and how the intervention timing might impact both the effectiveness of the intervention and the potential for misuse by exploring moderation prompts during typing and prior to sending (i.e., when the user presses ``send"). 


\section{Conclusion}\label{sec:conslusion}

While algorithmically-driven proactive moderation interventions that aim to reduce online toxicity are often well intentioned, our research points towards several ways in which they could be misused. We offer a critique of automated toxicity-detection systems, highlighting their potential for misuse by certain groups, for example through their potential to validate and gamify hateful attacks. 
We also highlight less explicit harms, including the potential for users to adapt their behaviours to conform to what they believe the algorithm considers to be~\textit{good}, resulting in \textit{algorithmic conformity} which could indirectly curtail freedoms of speech.
As such, we provide considerations for the designing of socio-technical systems that embed toxicity algorithms. Ultimately, we point to the need for truly multidisciplinary teams that work in interdisciplinary ways where algorithmic and HCI expertise work together to develop solutions which lessen potential negative impacts and misuse of these systems, while also improving the socio-technically situated positive and harm-reducing impacts that are potentially made possible with automated toxicity-detection algorithms when embedded meaningfully into socio-technical systems.

\section*{Acknowledgements}
This research was supported by UKRI through REPHRAIN (EP/V011189/1), the UK’s Research Centre on Privacy, Harm Reduction and Adversarial Influence Online.

\section*{Declaration of interest statement}
The authors report there are no competing interests to declare.

\section*{Data availability}
Anonymous workshop transcripts are available on request. Please Email the corresponding author. 

\bibliographystyle{elsarticle-harv}
\bibliography{referencesR3}

\appendix
\pagestyle{plain}
\setcounter{table}{0}
\captionsetup{labelformat=AppendixATables}

\begin{landscape}
\begin{longtable}{|>{\raggedright\arraybackslash}p{5cm}|>{\raggedright\arraybackslash}p{5cm}|>{\raggedright\arraybackslash}p{11cm}|}
\caption{Summary of Main Themes and Sub-Themes} \\
\hline
\textbf{Theme} & \textbf{Sub-theme} & \textbf{Summary} \\ \hline
\endfirsthead

\hline
\textbf{Theme} & \textbf{Sub-theme} & \textbf{Summary} \\ \hline
\endhead

\hline \multicolumn{3}{|r|}{\textit{Continued on next page}} \\ \hline
\endfoot

\hline
\endlastfoot

\textbf{Contextual factors impacting perceptions of toxicity and system effectiveness} 
& Conversational and interpersonal histories 
& The historical context of conversations, both personal and public, can influence the perception of toxicity. For example, language between friends may be considered toxic by an algorithm but not considered toxic by the individuals involved. \\ \cline{2-3} 

& Scale of abuse 
& The number of individuals involved in sending toxic messages and the audience size can amplify the harm caused. Messages part of network-level harassment~\citep{marwick2021morally} campaigns can be more harmful, even if individually they seem benign. \\ \cline{2-3} 

& Platform culture and affordances 
& Different platforms have distinct cultures and affordances, influencing the type and perception of toxic content. For instance, Twitter's algorithms could amplify and spread subtle forms of hate through trending topics. \\ \cline{2-3} 

& Social histories of oppression and power structures 
& Online content is influenced by offline socio-political power dynamics. Messages rooted in historical oppression (e.g., misogyny) need to be contextualised beyond their face value to understand their impact fully. \\ \hline

\textbf{End-user continuum of intention to send toxic messages} 
& The unaware and open to learning 
& Users who may unintentionally send toxic messages due to a lack of awareness. Proactive moderation can help educate and prevent future toxic behaviour. \\ \cline{2-3} 

& The emotionally triggered 
& Users who typically do not engage in toxic behaviour but may send harmful messages when emotionally triggered. These actions are often later regretted. \\ \cline{2-3} 

& Those using toxicity as a form of emotional arousal 
& Some users engage in toxic behaviour for emotional stimulation, such as fun or excitement. \\ \cline{2-3} 

& Those playing to an audience 
& Users who engage in toxic behaviour to gain attention or show off, especially in the presence of an audience. The presence of others can exacerbate their toxic actions. \\ \cline{2-3} 

& The determined and organised 
& Users with strong intentions to harm others, often part of coordinated efforts. These users are less likely to be deterred by moderation and may even adapt to circumvent it. \\ \hline

\textbf{Abuse and manipulation of systems with embedded toxicity models} 
& Validation 
& Toxicity scores provided by systems could unintentionally validate harmful behaviour, encouraging users to continue or escalate their actions. \\ \cline{2-3} 

& Gamification 
& Some users may treat toxicity scoring as a game, competing to achieve higher levels of toxicity or sharing their `scores' as a badge of honour. \\ \cline{2-3} 

& Model manipulation 
& Communities or users may attempt to manipulate toxicity detection models by providing false data, either to protect harmful content or to undermine the system's effectiveness. \\ \cline{2-3} 

& Circumvention 
& Users determined to send toxic messages may find ways to bypass the system, either by altering their language or using alternative methods of communication. \\ \hline

\label{tab:summary}
\end{longtable}
\end{landscape}

\end{document}